\documentclass[aps,prb,twocolumn,showpacs,amssymb]{revtex4}
\usepackage{bm}
\usepackage{graphicx}

\begin{document}

\title{de Haas-van Alphen effect versus Integer Quantum Hall effect}

\author{V.P. Mineev } 
\affiliation{Commissariat \`a l'Energie Atomique,
DSM/DRFMC/SPSMS 38054 Grenoble, France}

\date{\today}

\begin{abstract}
In frame of general stastical mechanics approach applied to 2D metal bar
we demonstrate the interrelationship between Landau diamagnetism, de Haas-van Alphen magnetization oscillations and the Integer Quantum Hall effect.    

\end{abstract}


\maketitle

\section{Introduction}

The goal of this article is to give a simple physical description of the Integer Quantum Hall effect in its relation to the Landau diamagnetism and the de Haas - van Alphen effect. It does not mean that we can give 
answer to all the questions concerning the problem.  In some respect we shall rather formulate the problems following from our treatment.
With this purpose we begin with repetition of  very instructive derivation of free electron gas diamagnetic moment due to
F.Bloch \cite{Bloch}, based on well known L.Landau calculations \cite{Landau} and taking into account the consideration given by
E.Teller \cite{Teller}. The important consequence of Bloch approach is a possibility to devide electrons
on two groups: (i) occupying the quantum states  in bulk  and (ii) in surface of metal.
Then we shall shaw that both groups are important in formation of Landau diamagnetic moment.
On the other hand only bulk electrons produce the oscillating magnetization that is the de Haas-van Alphen effect. Unlike to dHvA only surface electrons are responsible for quantized  Hall resistivity plateau and the existance at the same time of dissipationless longitudinal currents. 

In real situation it would be certainly quite naive to devide a heterojunction on bulk and surface regions.
The experiments clearly demonstrates that the currents in dissipativeless regime spread over whole specimen (see the review article \cite{Cage} and referencies therein). Moreover, the Integer Quantum Hall effect has been observed in the absence of edges that is on the samples of the Corbino geometry \cite{Dolg}.
In our opinion, however, this does not disprove the concept of division of electron states on bulk and edge states
but just  demonstrates that the sample in dissipativeless regime is devided on many channels or rivers
each of those contains
as electronic states localized far from the river banks and as well the bank states currying persistent currents. So, in our simplifyed treatment we shall work with completely homogeneous one channel model.

For the brevity we shall work with spinless fermions. The  spin degrees of freedom are easily included
in usual manner.  
We shall omit also an influence of disorder as unimportant for our model and leading  just to the appearance of Dingle factor in thermodynamic values.

\section{ Electron gaz in magnetic field:
role of bulk and surface electron states}

We shall discuss the 2D clean metal bar with length $A$ in $x$ direction and the width $B$ in $y$ direction $(|y|\le B/2)$ under perpendicular 
magnetic field ${\bf H}=(0,0,H)$. The thermodynamic potential of electron gas is
\begin{equation}
\Omega=-T\sum_\nu\ln\left(1+e^{\frac{\mu-\varepsilon_\nu}{T}} \right),
\end{equation}
where the electron energies $\varepsilon_\nu$ are determined as eigen values of Schroedinger equation
\begin{equation}
\left\{\frac{1}{2m}\left [\left(-i\frac{\partial}{\partial x}-eHy\right)^2 -\frac{\partial^2}{\partial y^2}\right ]+V(y)\right\}
\psi_\nu=\varepsilon_\nu\psi_\nu.
\end{equation}
We put $\hbar=c=1$ and $e=|e|$ throughout the paper. The potential $V(y)$ is negligible everywhere but near the edges where it grows up from zero at $|y|=B/2-\Delta$
to infinity at $|y|=B/2$. The length $\Delta$ is chosen much larger than magnetic length $\lambda_H=1/\sqrt{eH}$:
\begin{equation}
\frac{1}{(eH)^{1/2}}\ll\Delta\ll B
\end{equation}
The search of solution in the standart form
\begin{equation}
\psi_\nu=\exp(iqx)\varphi_\nu(y),
\end{equation}
where $q=2\pi Q/A$ and $Q$ is the integer,
leads us to the following eigen problem
\begin{eqnarray}
&
\hat H\varphi_\nu=\varepsilon_\nu\varphi_\nu,\nonumber\\
&\hat H=\frac{1}{2m}\left [\left(q-eHy\right)^2 -\frac{\partial^2}{\partial y^2}\right]+V(y).
\label{eSch}
\end{eqnarray}
It is clear that if the "equilibrium position"  $y_0=q/eH$ is limited by
\begin{equation}
-\frac{B}{2}+\Delta\le\frac{q}{eH}\le\frac{B}{2}-\Delta,
\label{eint}
\end{equation}
then to a quite good approximation the eigen functions of equation (\ref{eSch}) are Landau wave functions 
$$\varphi_{nq}(y)=\frac{1}{\sqrt{2^n n!\sqrt\pi\lambda_H}}\exp[-(y-y_0)^2/2\lambda_H^2]H_n[(y-y_o)/\lambda_H]$$
with $q$-independent eigenvalues
\begin{equation}
\varepsilon_n=\omega_c\left(n+\frac{1}{2}\right),
\label{eosc}
\end{equation}
$\omega_c=eH/m$. While if the $y_0$ is out the interval (\ref{eint}) then
the eigenvalues $\varepsilon_{\nu}=\varepsilon_{qn}$ are not so simple: they are $q$ dependent and tend to infinity when $|q|\to eHB/2$.

Applying the standart notations for quantum mechanical averages $\langle{\hat A}\rangle_\nu=\int dy(\varphi_\nu\hat A \varphi_\nu)$
we obtain the following equality: 
\begin{equation}
-H\frac{\partial \varepsilon_{nq}}{\partial H}=q\frac{\partial \varepsilon_{nq}}{\partial q} - \frac{(eH)^2}{m}\langle (y-y_0)^2\rangle_\nu,
\label{eequal}
\end{equation}
where
\begin{equation}
-\frac{\partial \varepsilon_{nq}}{\partial H}=-\left\langle \frac{\partial \hat H}{\partial H}\right\rangle_\nu=\langle \hat M_z\rangle_\nu,
\end{equation}
and
\begin{equation}
\frac{\partial \varepsilon_{nq}}{\partial q}=\left\langle\frac{\partial \hat H}{\partial q}\right\rangle_\nu=\langle \hat v_x\rangle_\nu.
\end{equation}

For the equlibrium value of the system magnetic moment at given temperature we have
\begin{eqnarray}
M=-\left (\frac{\partial \Omega}{\partial H}\right)_\mu=-\sum_{nQ}\frac{\frac{\partial \varepsilon_{nq}}{\partial H}}
{e^{\frac{\varepsilon_{nq}-\mu}{T}}+1}
\nonumber\\
=-\frac{A}{2\pi}\int\limits_{-eHB/2}^{eHB/2} dq\sum_{n=0}^{\infty}\frac{\frac{\partial \varepsilon_{nq}}{\partial H}}
{e^{\frac{\varepsilon_{nq}-\mu}{T}}+1}.
\end{eqnarray}
The integral over $q$  can be written as 
\begin{eqnarray}
&\int\limits_{-eHB/2}^{eHB/2} dq=\nonumber\\
&\left\{\int\limits_{-eH(B/2-\Delta)}^{eH(B/2-\Delta)}+\int\limits_{-\infty}^{-eH(B/2-\Delta)} +\int\limits_{eH(B/2-\Delta)}^{\infty}\right\}dq, 
\end{eqnarray}
where the infinite limits are taken due to fast exponential convergency of integral when $\varepsilon_{nq}\to \infty $ at $q$ outside the interval
(\ref{eint}). Correspondingly the magnetic moment  presents the sum of three terms
\begin{equation}
M=M_1+M_2+M_3
\end{equation}
For the first term the energy levels have $q$ independent value (\ref{eosc}), hence
\begin{equation}
M_1=-\frac{eS}{2\pi}\sum_{n=0}^{\infty}\frac{\varepsilon_{n}}
{e^{\frac{\varepsilon_{n}-\mu}{T}}+1}
\end{equation}
where $S=AB$ is the bar area.
For the second term, by making use the equality (\ref{eequal}) and omiting the contribution from $\propto \langle(y-y_0)^2\rangle_\nu$
which is of the order of $\Delta/B$ in comparison with other terms we obtain
\begin{eqnarray}
&M_2=\frac{A}{2\pi H}\int\limits_{-\infty}^{-eH(B/2-\Delta)}qdq\sum\limits_{n=0}^{\infty}\frac{\frac{\partial \varepsilon_{nq}}{\partial q}}
{e^{\frac{\varepsilon_{nq}-\mu}{T}}+1}\nonumber\\
&=-\frac{eS}{4\pi }\int\limits_{-\infty}^{-eH(B/2-\Delta)}dq\sum\limits_{n=0}^{\infty}\frac{\frac{\partial \varepsilon_{nq}}{\partial q}}
{e^{\frac{\varepsilon_{nq}-\mu}{T}}+1}\nonumber\\
&=\frac{eST}{4\pi }\sum\limits_{n=0}^{\infty}\ln\left(1+e^{\frac{\mu-\varepsilon_n}{T}} \right).
\end{eqnarray}
Taking into account that $M_2=M_3$ finally we have
\begin{equation}
M=M_1+2M_2=
-\left(\frac{\partial \Omega}{\partial H}\right)_\mu,
\end{equation}
where
\begin{equation}
\Omega=-\frac{eHST}{2\pi}\sum\limits_{n=0}^{\infty}
\ln\left(1+e^{\frac{\mu-\varepsilon_n}{T}} \right).
\end{equation}

One can rewrite this result also as 
\begin{eqnarray}
&M=M_1+2M_2=\left(H\frac{\partial}{\partial H}+1\right)\left(-\frac{\Omega}{H}\right)_\mu=\nonumber\\
&=\left(H\frac{\partial}{\partial H}+1 \right)\frac{eST}{2\pi}\sum\limits_{n=0}^{\infty}
\ln\left(1+e^{\frac{\mu-\varepsilon_n}{T}} \right).
\end{eqnarray}
As follows from the derivation the first term here  $M_1=-H\partial/\partial H(\Omega/H)_\mu$ 
caused by electrons occupying the Landau states
situated in bulk of metal. The second term $2M_2=-\Omega/H$ is due to the electrons filling the edge states.
Let us look now what roles play these two group of electrons in observable physical effects.

\section{Landau diamagnetism}

In  low field limit $\omega_c\ll T$ the application of Euler-Maclaurin summation formula \cite{LL}
yields
\begin{eqnarray}
&M=\left(H\frac{\partial}{\partial H}+1 \right)
\left\{-\frac{\Omega_{H=0}}{H}\right.\nonumber\\
&\left.\left.+\frac{eST}{2\pi}\frac{eH}{24m}{\partial 
\ln\left(1+e^{\frac{\mu-\varepsilon}{T}}\right)}\right.\left/{\partial \varepsilon}\right |_{\varepsilon=0} \right\}
=-\frac{e^2HS}{24\pi m}.
\end{eqnarray}
We see that both group of electrons give the equal contribution to Landau diamagnetism. It is worth to be noting that half of this momentum is the sum of orbital magnetic moments of electronic states in the bulk of material. And another half
is associated with the persistent current curried by electrons occupying the orbits skipping along the specimen surface. This persistent current is similar to the persistent currents in mesoscopic rings (see for instance paper \cite{CRG}) and has pure single particle nature unlike to superconducting or superfluid currents which are persistent due to multiparticle coherence.

\section{de Haas-van Alphen effect}

In high field limit  $\omega_c\gg T$ the application of Poisson summation formula
 \cite{LL} yields 
\begin{eqnarray}
&M=\left(H\frac{\partial}{\partial H}+1 \right) \left\{-\frac{\Omega_{H=0}}{H}+\right.\nonumber\\
&\left.+\frac{eST}{2\pi}2\Re\sum\limits_1^\infty\int\limits_0^\infty dx
\ln\left(1+e^{\frac{\mu-\omega(x+1/2)}{T}}\right)e^{2\pi ilx}\right\}=\nonumber\\ 
&=\left(H\frac{\partial}{\partial H}+1 \right)\left\{-\frac{eST}{2\pi}\sum\limits_{l=1}^\infty
\frac{(-1)^l}{l}\frac{\cos(2\pi l\mu/\omega_c)}{\sinh(2\pi^2lT/\omega_c)}\right\}
\end{eqnarray}
The expression in paranthesis $\{....\}=-\Omega_{osc}^{2D}/H$ where 
$\Omega_{osc}^{2D}$  coincides exactly with
found in the paper \cite{ChMin},
where also the spin splitting and the impurity scattering have been taken into account.
To obtain the oscillating part of magnetization, that is the de Haas - van Alphen effect,
one must differentiate the fast oscillating $\cos(2\pi l\mu/\omega_c)$. 
\begin{equation}
M_{osc}=
\frac{eST\mu}{\omega_c}\sum\limits_{l=1}^\infty
(-1)^{l+1}\frac{\sin(2\pi l\mu/\omega_c)}{\sinh(2\pi^2lT/\omega_c)}
\label{emu}
\end{equation}
Hence, it is clear that contrary to Landau diamagnetism the only bulk electrons are responsible for de Haas - van Alphen signal.

Unlike to 3D metal the chemical potential in eqn (\ref{emu}) is strongly oscillating function of magnetic field. To find it, following the papers  \cite{ChMin,Cham,Grig}, we calculate the number of particles
by means of thermodynamic relation
\begin{equation}
N=-\left(\frac{\partial \Omega}{\partial\mu}\right)_T
\end{equation}
and then resolve this equation in respect to chemical potential.
The result is as follows
\begin{equation}
\mu=\varepsilon_F+\mu_{osc},
\label{e23}
\end{equation}
where
\begin{equation}
\mu_{osc}=\frac{H}{N}M_{osc}
\label{e24}
\end{equation}
So, the oscillating behavior of magnetization and the chemical potential are determined self-consistently in accordance with written above equations (for more details see \cite{ChMin,Cham,Grig}). The experimental evidence of de Haas-van Alphen oscillations in 2D heterostructures has been established recently \cite{Wilde}.     We should stress here that both magnetization and the chemical potential oscillations are determined by bulk electron states.

\section{Integer Quantum Hall effect}

In presence of  a current in $x$-direction there is Lorentz force shifting the charge carriers in $y$-direction to the bar edges of where the extra (or lack of) charge appears leading to Lorenz force compensation by Coulomb interaction.  As result  the values of chemical potential at the opposite banks of bar differ each other by the Hall voltage
\begin{equation}
\mu(B/2)-\mu(-B/2)=eU_H. 
\label{ech}
\end{equation}
The local current density is 
\begin{equation}
j_x=\frac{\partial {\cal M}_z}{\partial y},
\end{equation}
where ${\cal M}_z=M_z/S$ is the magnetic moment density.
Hence the current is given by  
\begin{equation}
J_x=\int
\limits_{-B/2}^{B/2} j_xdy={\cal M}_z(B/2)-{\cal M}_z(-B/2),
\end{equation}
and as the magnetic moment one must  take the surface part of magnetic moment.  Thus
\begin{eqnarray}
&J_x=\frac{eT}{2\pi}\sum\limits_{n=0}^{\infty}\left\{\ln \left(1+\exp{\frac{\mu(B/2)-
\varepsilon_n}{T}}\right)\right.\nonumber\\
&\left.-\ln \left(1+\exp{\frac{\mu(-B/2)-\varepsilon_n}{T}}\right)\right\}.
\end{eqnarray}
This current is not accompanied by a dissipation, becouse edges of specimen are at  constant potential:
the values of chemical potential in eqn (\ref{ech}) are not $x$-dependent. So this current is quite similar to described above persistent edge currents responsible for Landau diamagnetism. Unlike to the latter
the currents at the opposite edges are not equal due to chemical potential difference.

At small currents and, hence, at small Hall tensions we have for  Hall conductance
\begin{equation}
G_{xy}=\frac{J_x}{U_H}\simeq\frac{e^2}{2\pi}\sum\limits_{n=0}^{\infty}
\frac{1}{e^{\frac{\varepsilon_n-\mu_s}{T}}+1}.
\label{eIQH}
\end{equation}
So, for negligibly small $U_H$ the Hall conductance at low temperatures has quatized values determined by number of Landau levels below the chemical potential. 

This property takes place if 
the surface chemical potential 
$\mu(x,\pm B/2)=\mu_s$ does not oscillate with magnetic field dissimilar to the chemical potential in the bulk where it  is strongly oscillating function according to  equations (\ref{e23}), (\ref{e24}).  
Nonoscillating behavior of chemical potential is typical for 2D electron system connected with reservoir
 \cite{Cham}. Here we can expect that the chemical potential of ensemble of the surface electronic states is maintained at the constant value by the reservoir of electronic states in the bulk.
The equilibrium between the surface and the bulk electron subsystem is supported
 by means of electric potential of changing in space electron density \cite{Finkel}. 
  
 %The bulk and the surface chemical potentials coincides only at magnetic fields corresponding to completely filled Landau levels.  

%At fixed  value of current  the Hall conductance quantization is provided by the quantization of Hall voltage. The latter  is supported by  absence of charge space  rearrangement near the edges. So, the existence of Hall conductance plateaus follows from the stability of edge charge for the magnetic fields near the filled Landau levels. The mechanism of this stability is related to formation of "bubbles"
%empty of electrons filling of slightly underfilled Landau level. The bubbles are separated from surrounding 2D electron gas by an incompressible stripes of integer Landau level filling  \cite{Finkel}.
%The stability is obviously violated  in narrow fields 
%intervals corresponding to approximate coincidence of Fermi energy with energy of the upper half filled Landau level, where many Landau orbits 
%are not occupied and
%the Pauli principle do not prevent to local changes of electron density.

%The Hall voltage quantization should disappear at small enough fields when the the edge charge screening length starts to be comparable with magnetic length $\lambda_H=1/\sqrt{eH}$.

\section{Conclusion}

The zero temperature derivation of edge currents and the Quantized Hall effect relationship
has been introduced in the paper \cite{McS}. Recently the general thermodynamic arguments were used
for the description of Hall effect in terms of diamagnetic currents \cite{Streda}. 

Besides the statistical mechanics treatment of the IQHE as 
an equilibrium   phenomenon at finite temperature and fixed number of particles here we have pointed out the relationship between
all type of oscillation phenomena in 2D metals.  The difference of the surface and the bulk chemical potentials comes out as an inevitable property of our approach. The similar physical conjecture has been put forward and qualitatively described by V.Egorov \cite{E}.

The currents in the field interval of Hall plateau are persistent currents similar to those responsible for Landau diamagnetism. Their stability is provided by sort
of energetic bariers preventing the dissipative electron density redistribution near filled Landau level \cite{Finkel}. On the conrary, the dissipative regime arises in 
the magnetic field intervals near the half filled Landau levels when the freedom for the electron motion is not limited by the Pauli principle.

By different approach the formula (\ref{eIQH}) is derived  recently  by Champel and Florens \cite{CF}.
We should stress however that these authors do not distinguish the surface and bulk chemical potential
values. The absence of oscillating part of the chemical potential is provided by presence of  smooth in space but strong in amplitude potential disorder. However, the disorder potential at the specimen edges is taken as $x$-coordinate independent.  As we already mentioned the experimental evidence of strong oscillations of magnetic moment (related with the bulk chemical potential oscillations)  in high mobility heterostructuters \cite{Wilde} have been reported recently.

\smallskip

{\bf {Acknowledgments}}

\smallskip

It is my pleasure to express the gratitude to V.S.Egorov and T.Champel for countless enlightening discussions.

\end{document}